\documentclass[12pt,3p]{elsarticlem2}
\usepackage{hyperref,times,pifont,graphicx,amsmath,amssymb,lineno,array}
\biboptions{numbers,comma,square,sort&compress} 
\bibpunct{[}{]}{,}{n}{,}{,}
\usepackage{lineno}
\usepackage{color}
\usepackage[footnotesize,bf]{caption}
\renewcommand{\d}{\text{d}}
\newcommand{\PreserveBackslash}[1]{\let\temp=\\#1\let\\=\temp}
 \let\PBS=\PreserveBackslash

\date{}

\begin{document}

\begin{frontmatter}

\title{Hydrogen peroxide thermochemical oscillator as driver for primordial RNA replication} 
%short title: H$_2$O$_2$ oscillator drives RNA replication
\author{ Rowena Ball}
\ead{Rowena.Ball@anu.edu.au}
\address{Mathematical Sciences Institute, The Australian National University,  Canberra 0200, Australia}

\author{John Brindley}
\address{School of Mathematics, University of Leeds, Leeds LS2 9JT, U. K. }

\begin{abstract}
This paper presents and tests a previously unrecognised mechanism for driving a replicating molecular system on the prebiotic earth. It is proposed that cell-free RNA replication in the primordial soup may have been driven by self-sustained oscillatory thermochemical reactions. To test this hypothesis a well-characterised hydrogen peroxide oscillator was chosen as the driver and complementary RNA strands with known association and melting kinetics were used as the substrate. An open flow system model for the self-consistent, coupled evolution of the temperature and concentrations in a simple autocatalytic scheme is solved numerically, and it is shown that thermochemical cycling drives replication of the RNA strands. For the (justifiably realistic) values of parameters chosen for the simulated example system,  the mean amount of replicant produced at steady state is 6.56 times the input amount, given a constant supply of substrate species.  The spontaneous onset of sustained thermochemical oscillations via slowly drifting parameters is demonstrated, and a scheme is given for prebiotic production of complementary RNA strands on rock surfaces. 
\end{abstract}


\end{frontmatter} 
\newpage

 \section{Introduction}
 
Molecular self-replication is a fundamental process of all living organisms, and its initiation in the environment provided by the early Earth must have been essential for the emergence of life itself.   In modern cells DNA replication is accomplished  isothermally at moderate temperatures by a working army of enzymes using the free energy of ATP hydrolysis, and the strongly driven nonequilibrium condition is maintained by respiration and ion transport across cell membranes. However in the primordial soup there were no cells and no  protein enzymes.  
The `RNA world'  hypothesis holds that cell-free RNA communities grew on solid surfaces and replicated, before the evolution of DNA 
 \cite{Neveu:2013}.  What energy source may have driven RNA replication  in such an environment?
 
 It does seem that thermal cycling  would be required to drive RNA replication in the absence of cellular (or other) machinery, because heat is required to dissociate double-stranded or multiplex RNA and a cooler phase is necessary for replication and annealing, given a supply of substrate template. This fact  is often overlooked in hypotheses about the origin of life, as pointed out by Kov\'a\u{c} et al. \cite{Kovac:2003}.   Exponential replication of RNA duplexes, fed by hairpins derived from an alanine tRNA and driven by externally imposed thermal cycling between 10 and 40$^\circ$C, was achieved by Krammer et al.  \cite{Krammer:2012}. They  proposed that, in the primordial soup, thermal  cycling  may have been provided by laminar convection between hot and cold regions in millimetre-sized rock pores \cite{Braun:2003}, and that the substrate oligonucleotides could be concentrated by thermophoretic trapping. 

It has been suggested by Matatov \cite{Matatov:1980} that chemical energy  also may have played a role in prebiotic evolution, because, in the laboratory, heat liberated in the decomposition of aqueous hydrogen peroxide (H$_2$O$_2$)  supported  formation of amino acids from simpler precursors. 
In this work we propose that a natural mechanism for driving self-sustained thermal cycling is a thermochemical oscillator, and demonstrate in a specific case that the temperature cycling for driving the amplification of template RNA in environments where early life may have evolved can be provided by a thermochemical oscillator driven by exothermic reactions of  H$_2$O$_2$. 
 
Exothermic reactions of aqueous H$_2$O$_2$  are well-known to give rise to robust, self-sustained  thermochemical oscillations with frequencies around 0.02--0.005\,s$^{-1}$   \cite{Ramirez:1969,Chang:1975,Wirges:1980,Zeyer:1999}.  As our specific example we have chosen, with some justification expressed below, the oxidation of  thiosulfate ion (S$_2$O$_3^{2-}$) by H$_2$O$_2$, which is fast and highly exothermic:
 \begin{equation}\tag{R1}
 \rm{Na}_2\rm{S}_2\rm{O}_3 + 2\rm{H}_2\rm{O}_2 \rightarrow \frac{1}{2}  \rm{Na}_2\rm{S}_3\rm{O}_6 + \frac{1}{2}\rm{Na}_2\rm{S}\rm{O}_4 + 2\rm{H}_2\rm{O}.
 \end{equation}
 The thermokinetics and thermochemistry of (R1) have been well-studied, and  experimental data from the literature are collected in table \ref{thiosulfate}. The reaction is first order in both reactants.
 
 Oscillatory thermoconversion is typical of highly energetic, thermally sensitive, liquids such many of the peroxides \cite{Ball:2013}. The physical basis  is as follows: In such liquids (and also in some vapor and gas phase substances), which have high specific heat capacity, the heat of reaction can be absorbed by the many intermolecular vibrational modes and intramolecular rotational modes. So the temperature rises but slowly as reaction proceeds, until these modes become saturated and consequently the temperature spikes --- the heat cannot be absorbed any more by quantized excitations. But locally the reactant is depleted, so the temperature falls to a minimum before reactant accumulation allows reaction to occur and heat to be released, and the cycle begins again. With a steady supply of reactant these self-sustained thermal oscillations can continue indefinitely.
 
    \begin{table}[t]\caption{\label{thiosulfate} Experimental data for reaction (R1), presented in chronological order. } 
\footnotesize\centerline{
 \begin{tabular}{p{0.24\columnwidth}p{0.2\columnwidth}p{0.24\columnwidth}p{0.08\columnwidth}}
 \hline\hline
 $A$ (l\,mol$^{-1}$s$^{-1}$) &  $E$ (kJ/mol) &    $-\Delta H$ (kJ/mol) & Ref. \\
 \hline\hline
$6.85 \times 10^{11} $ & 76.52 			& 573.0		& \cite{Cohen:1962}\\
$ 2.13 \times 10^{10}$   & 68.20			& 585.8		& \cite{Lo:1972}\\  
$ 7.33 \times 10^{11}$  & 78.24			& 594.1		& \cite{Williams:1974} \\
$1.63 \times 10^{10}$   & 68.12			& 612.5		& \cite{Chang:1975}\\ 
$6.8 \times 10^{11}$      & 77.0				&not measured			&\cite{Guha:1975}\\
$2.00\times 10^{10}$    & 68.20			& 586.2		&\cite{Lin:1981}\\
$2.0  \times 10 ^{10}$    & 68.3				&  562.8		&\cite{Grau:2000}\\
 \hline\hline
 \end{tabular}
 }
 \end{table}

\subsection{H$_2$O$_2$ and thiosulfate on the early earth}

One school of thought holds that the geochemical environment for the emergence of life was provided by 
submarine hydrothermal systems and hot springs 
\citep{Tang:2007}. Experiments reported by Foustoukos et al. \citep{Foustoukos:2011} strongly support  the conjecture that H$_2$O$_2$ is produced near hydrothermal vents when oxygenated seawater mixes with vent fluid.  

Another source of  H$_2$O$_2$ production in such an environment involves a surface reaction of  pyrite (FeS$_2$)  with H$_2$O  and in experiments  0.391--0.567\,mM/m$^2$  was produced \cite{Borda:2001,Borda:2003}. The specific surface area of pyrite is 2.0--4.0\,m$^2$/g \citep{Pugh:1981}, so there is a very real possibility that high concentrations of H$_2$O$_2$ could build up locally and be supplied at a constant rate for long periods of time.  A very credible body of work (see references in \cite{Borda:2001}) holds  that the most primitive photosynthetic cells used H$_2$O$_2$ as an electron donor, so it is reasonable to assume that H$_2$O$_2$ was produced long before those organisms evolved. 
Moreover it  is believed that pyrite was present on the early earth. On the modern earth, pyrite deposits associated with hydrothermal activity  can reach thicknesses of tens to hundreds of meters and spread over thousands of square kilometers within the crust \cite{Barrie:1999}. 

Prebiotic production of H$_2$O$_2$ also may have  occurred by photochemical disproportionation of the superoxide radical  (O$_2^{-}$)   in sunlit waters to H$_2$O$_2$  \citep{Bar:2009}. 

Thiosulfate ion  is observed to occur  in  hydrothermal waters of Yellowstone National Park \citep{Xu:1998}, so it is reasonable to conjecture that it was present in hydrothermal environments on the early earth. 
  
Porous rocks around hydrothermal vents therefore    could  provide  microenvironments   where naturally self-sustaining thermochemical oscillations may be set up --- namely, a localized, nonequilibrium forced flow system, and a supply of H$_2$O$_2$ and thiosulfate ion. If  a supply of RNA oligonucleotides, perhaps synthesized on and dislodged from pore surfaces \citep{Kiedrowski:2001}, is added to this recipe for primordial soup, replication may be driven by  thermochemical temperature cycling.  
  
\section{Model, data and methods}
For the reactive system in a single rock pore we employ a spatially homogeneous flow model; in other words, a continuous stirred tank reactor  (CSTR) paradigm. An \textit{a posteriori} assessment of the rationale and validity of this model is given in  section \ref{section4}.1.  The following dynamical system models the coupled  evolution of the reactant concentrations in (R1) and the temperature:
\begin{align}
V\frac{\d c_\text{v}}{\d t} &= -Vk_1(T)c_\text{v}c_\text{w} +F(c_\text{v,f}-c_\text{v})\label{e1}\\
V\frac{\d c_\text{w}}{\d t} &= -V2k_1(T)c_\text{v}c_\text{w} +F(c_\text{w,f}-c_\text{w})\label{e2}\\
V\bar{C}\frac{\d T}{\d t}& = (-\Delta H_1)Vk_1(T)c_\text{v}c_\text{w} \notag \\
&\hspace{15mm} -F\bar{C}(T-T_\text{f}) - L(T-T_\text{a}), \label{e3}
 \end{align}
where the reaction rate constant  $k_1(T) = A_1\exp(-E_1/(RT))$, $c_\text{v} $ is the concentration of thiosulfate and $c_\text{w} $ is the concentration of H$_2$O$_2$. The symbols and notation used in these and  following equations are defined in table \ref{table1}. 
\begin{figure}[t]
\centerline{
\includegraphics[scale=1]{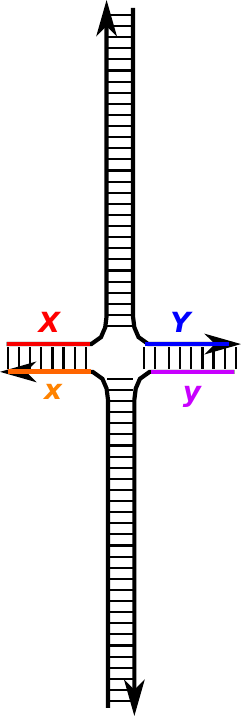}
}
\caption{\label{figure1} The quadruplex RNA species $Z$. The duplex replicator species are $XY$ and $xy$, which for the purposes of the simulation are equivalent. See ref. \cite{Krammer:2012} for more details. (Online version in colour.)}
\end{figure} 

We  used the following minimum subset of reactions from ref. \cite{Krammer:2012}  that can produce duplex RNA by autocatalysis:
 \begin{align}\tag{R2}
 X + Y  + XY &\overset{k_2}\longrightarrow Z\\
 Z &\overset{k_3} \longrightarrow 2XY, \tag{R3}
  \end{align}
where $X$ and $Y$ represent  RNA complementary single-strands, the RNA duplex $XY$ is the replicator and $Z$ is the RNA quadruplex, the form of which is cartooned in figure \ref{figure1}. In the reaction volume these species evolve as
 \begin{align}
& V\frac{\d c_\text{x+y}}{\d t} = -V2k_2(T)c_\text{x}c_\text{y}c_\text{xy} + F(c_\text{x,f+y,f}-c_\text{x+y})\label{e4}\\
&V\frac{\d c_\text{xy}}{\d t} = -Vk_2(T)c_\text{x}c_\text{y}c_\text{xy}  + V2k_3(T) c_\text{z} +F(c_\text{f,xy}-c_\text{xy})\label{e5}\\
& V\frac{\d c_\text{z}}{\d t} = Vk_2(T)c_\text{x}c_\text{y}c_\text{xy} - Vk_3(T) c_\text{z} -Fc_\text{z}, \label{e6}
 \end{align}
 where the notations $c_\text{x+y}$ and $c_\text{x,f+y,f}$ are shorthand for $c_\text{x}+c_\text{y}$ and $c_\text{x,f} + c_\text{y,f}$,  $k_2(T)= A_2\exp(-E_2/(RT))$ and $k_3(T)= A_3\exp(-E_3/(RT))$. (These are to be considered as empirical rate constants; (R2) as written is not intended to imply that an elementary three-body collision occurs.)

  The enthalpy balance includes contributions from all reactions: 
\begin{multline}\label{e7}  
 V\bar{C}\frac{\d T}{\d t} = (-\Delta H_1)Vk_1(T)c_\text{v}c_\text{w} \\
                                                     +  (-\Delta H_2)Vk_2(T)c_\text{x}c_\text{y}c_\text{xy}
                                                     +(-\Delta H_3)Vk_3(T)c_\text{z}   \\
                                                     - F\bar{C}(T-T_\text{f}) - L(T-T_\text{a}). 
\end{multline}
We  derived the activation energies, pre-exponential factors, and reaction enthalpies from the rate data in  \cite{Krammer:2012}, see the Appendix. 
Equations \eqref{e1}--\eqref{e3} (no RNA supply) and equations \eqref{e1}--\eqref{e2} and \eqref{e4}--\eqref{e7} (RNA single-strands and duplexes supplied), were integrated   using a stiff integrator from reasonable initial conditions. The stability of steady state solutions was assessed over a range of the inflow temperature, $T_\text{f}$, by solving the corresponding eigenvalue problem  and flagging points where an eigenvalue changed sign.  These bifurcation points then were followed over a range of the thermal conductance $L$ to obtain stability maps.  Numerical values of the fixed parameters are given in table \ref{table1} and  are discussed further in~ the Appendix.

\section{Results}
\begin{figure}[t]
\centerline{
\includegraphics[scale=1]{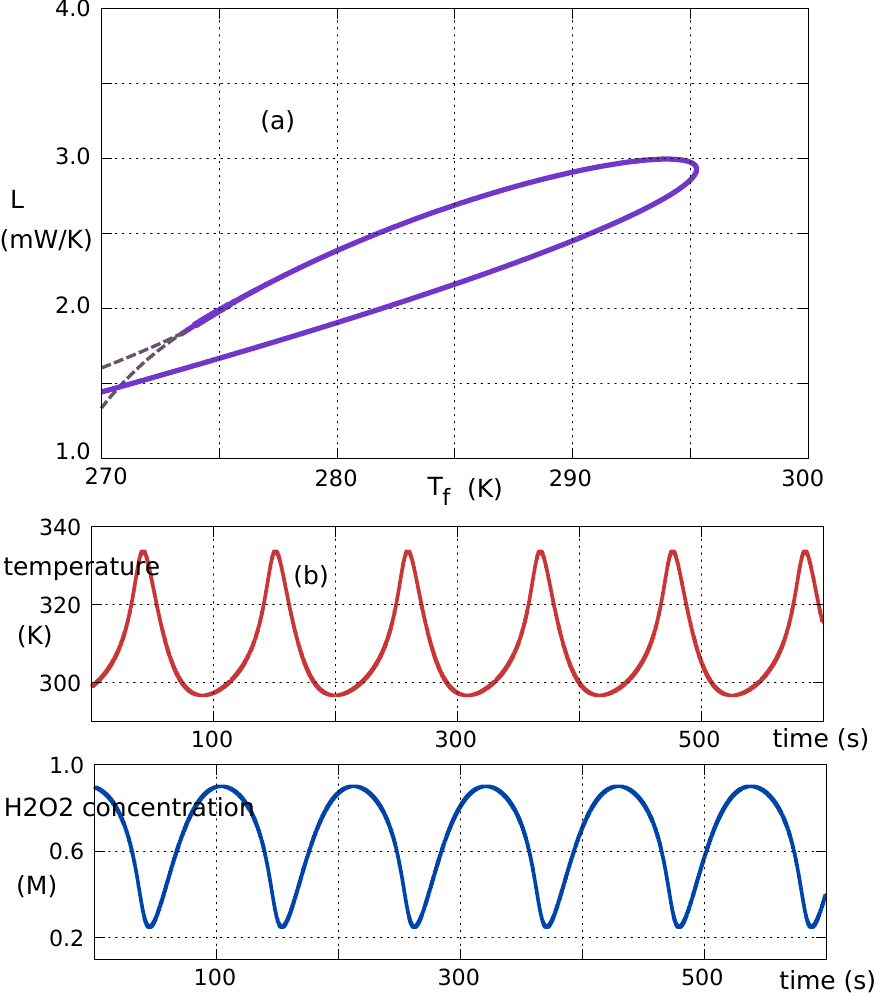}
}
\caption{\label{figure2} Computed data from equations  \eqref{e1}--\eqref{e3}: (a) Hopf (solid  line) and saddle-node (dashed  line) bifurcation loci; (b) Time series for $T_\text{f}=282$\,K, $L=2.4$\,mW;   the oscillation period  is  108.7\,s.  (Online version in colour.)}
% Aspect ratio = 0.880, word estimate = 190
\end{figure}

First we examined the behavior of the standalone H$_2$O$_2$/thiosulfate system. Figure \ref{figure2}(a) shows the computed stability map. 
The r\'egime of threefold multiplicity mapped by the locus of saddle-node bifurcations is included for completeness, but this r\'egime is largely irrelevant  for our purposes since the temperature  is too low to drive RNA replication.  At each point within the Hopf bifurcation loop the stable solution is a limit cycle. The time series for a selected point in the loop is shown in figure \ref{figure2}(b). 

\begin{figure}[t]
\centerline{
\includegraphics[scale=1]{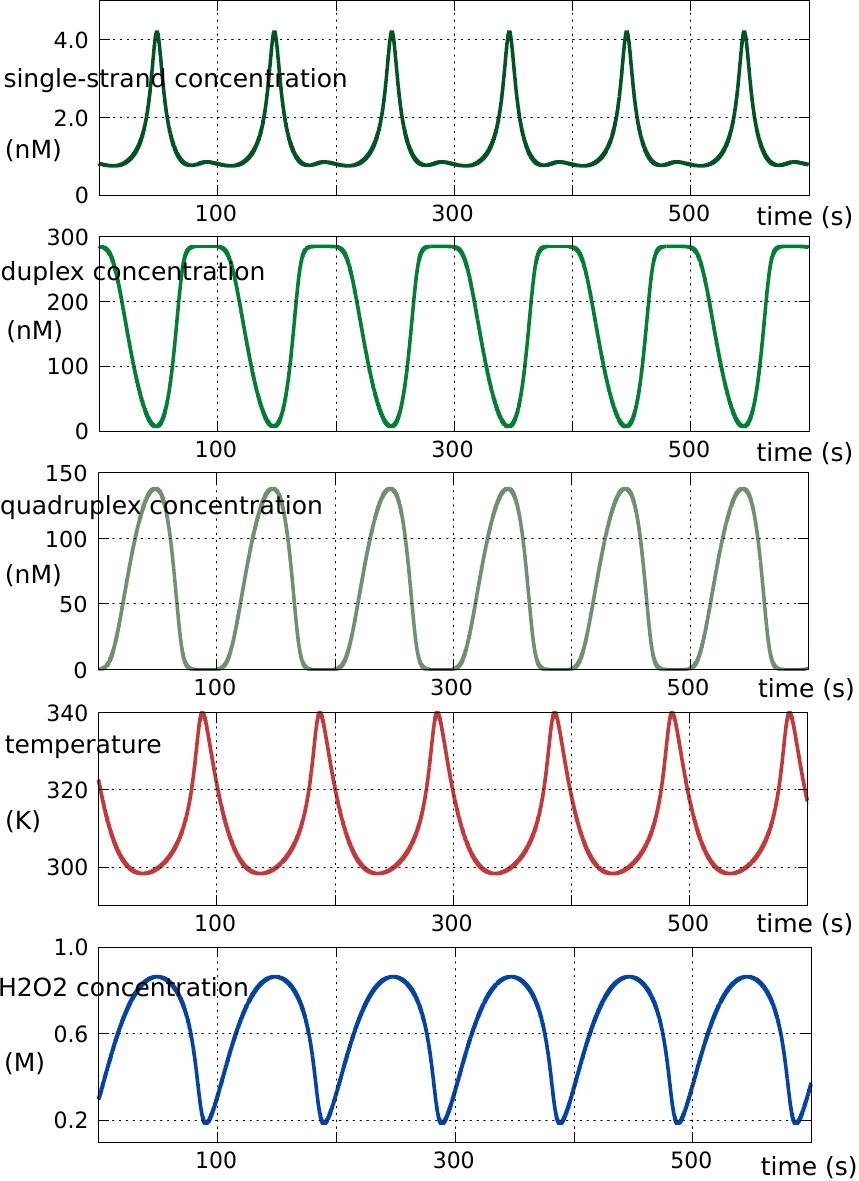}
}
\caption{\label{figure3}Time series from equations \eqref{e1}, \eqref{e2} and \eqref{e4}--\eqref{e7} for $T_\text{f}=282$\,K, $L=2.3$\,mW;  period of the oscillation is  99.3\,s.  (Online version in colour.)}
% Aspect ratio = 0.729, word estimate = 226
\end{figure}

When reactions (R2) and (R3) are included, the  stability map and temperature time series are  indistinguishable numerically from those in figure \ref{figure2} because 
the contributions of the second and third terms on the right hand side of equation \eqref{e7} are  relatively insignificant.

However, time series data rendered in figure \ref{figure3} for the concentrations of single-strands, duplex, and quadruplex show that the H$_2$O$_2$/thiosulfate thermochemical oscillator can indeed drive the RNA replication process. This time series  was computed for a slightly lower value of  $L$  and has a shorter period and higher temperature amplitude than that in figure \ref{figure2}(b).

The period  is determined by the  ratio $\phi\equiv (\bar{C} E_1)/(R c_\text{w,f}(-\Delta H_1))$, and in general a large value of the numerator (high specific heat and activation energy) corresponds to longer cycling times and a large value of the denominator (high specific reaction enthalpy) corresponds to shorter period oscillations.  The physical parameters $ V/F$ (mean residence time), $T_\mathrm{f}$, $T_\mathrm{a}$, and $L$ also affect the period, as might be inferred from the locus of Hopf bifurcations in figure   \ref{figure2}(a) \cite{Gray:1994}. The hydrogen peroxide oscillator has a variable period of the right order --- around 80 to 110 seconds --- to drive the replication of small RNAs. If the period is too long the RNAs may decay faster than replication can amplify them. If the period is too short the strands do not separate completely and replication may fail.

But the period  is not tuned to the putative doubling time for replicant concentration as was the thermal cycling time in \cite{Krammer:2012}.  As our system is an open flow system operating at dynamic steady state, where it has `forgotten' the initial conditions, the doubling time is a different quantity from that for the closed batch system in \cite{Krammer:2012}. Here it is defined as a time interval during which  the cumulative replicant concentration at the outlet, $c_\text{xy}+c_\text{z}$, becomes  equal to twice the inflow concentration $c_\text{xy,f}$.  In equations \eqref{e1}, \eqref{e2} and \eqref{e4}--\eqref{e7} the variables are evolved self-consistently and the doubling time is temperature-dependent, and therefore changes continuously. We can make use of average quantities though.

From computed datafile, the replicant concentration $c_\text{xy}+c_\text{z}$ integrated over one period of~99.3\,s (using the trapezoidal rule on 50,000 data points for high accuracy) is  $2.281 \times 10^{-5}$\,mol\,s/l, which yields an instantaneous average concentration  $\bar{c}_\text{xy}+\bar{c}_\text{z}=2.297\times 10^{-7}$\,mol/l and  an average doubling time of 0.31\,s, for $c_\text{xy,f}=3.5\times 10^{-8}$\,M.  This does not infer that replication is exponential, it simply means that on average it takes 0.31\,s for the cumulative amount of replicant produced to become equal to twice the feed amount. 

Another measure of the system's performance is the ratio $ \bar{c}_\text{xy}+\bar{c}_\text{z}/c_\text{xy,f}= 6.56$. This means the system has been designed so that in steady flow state, for the ideal case where no decay of replicant occurs within the reaction volume, the mean outlet replicant amount is 6.56 times the inlet duplex amount.

 Let us consider the phase relationships in figure \ref{figure3} over one period.
 Single-strands are consumed as the temperature rises  because duplex production by (R3), which requires heat, increases.  Correspondingly the concentration of quadruplex falls, almost to zero near the temperature maximum.  Single-strands  begin to accumulate, but there is only a small bump before quadruplex production  picks up as the temperature declines. The presence of this  bump  means that the oscillation is actually quasiperiodic, and reminds us that the dynamical system of equations \eqref{e1}--\eqref{e2} and \eqref{e4}--\eqref{e7}  is capable of complex periodic and even chaotic behaviour. This may lend additional, powerful capabilities to a molecular replicating system.  For example, biperiodic temperature response is capable of replicating two different RNA species, and nature may well have done exactly that in the primordial rock pores. Duplex concentration attains a broad maximum, then declines in phase with the temperature as reaction (R2) is favoured. The inflow delivers more single-strands, which accumulate even though they allow more production of quadruplex. 

\subsection{Development of self-sustained oscillations}

\begin{figure}
\centerline{
\includegraphics[scale=1]{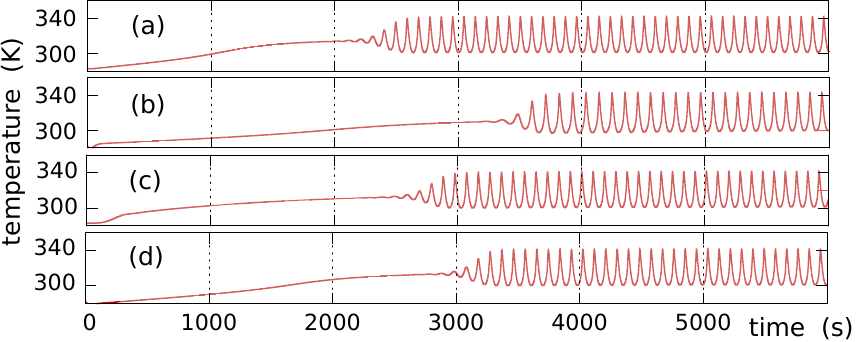}
}
\caption{\label{figure4} Onset of oscillations via slow drift of parameters. (a) $p\equiv c_\text{w,f}$, (b) $p\equiv T_\text{a}$, (c) $p\equiv F$, (d) both $p_1\equiv c_\text{w,f}$ and $p_2\equiv T_\text{a}$.  (Online version in colour.)}
% Aspect ratio = 2.48, word count = 81
\end{figure}

We have addressed also the question of how self-sustained H$_2$O$_2$ thermochemical oscillations may have arisen spontaneously in rock pores on the early earth. The fuzzy answer is that parameters may drift quasistatically into an oscillatory r\'egime, such as the loop of Hopf bifurcations in figure \ref{figure2}(a). 
A precise answer is provided by simulating various scenarios where one or more tunable parameters $p$ in equations \ref{e1}--\ref{e3} is given the following increasing  time dependence, such that $p$ approaches a constant value:
\begin{equation}\label{p}
p=p_\infty\left(1-\alpha \exp\left(-\gamma t\right)\right). 
\end{equation}
A time series from equations \eqref{e1}--\eqref{e3} using \eqref{p} for one or more parameters is expected to show the temperature  increasing slowly, then beginning to wobble, then settling into self-sustained oscillations.  Some examples are shown in figure \ref{figure4}. Of course, such evolutions could proceed for millions of years before the onset of periodic behaviour, but we we have set the time constant $\gamma$ in equation~\eqref{p} more conveniently. 

\section{Discussion\label{section4}}
\subsection{Validity and relevance of CSTR paradigm}
The CSTR paradigm described by equations  \eqref{e1}--\eqref{e2} and \eqref{e4}--\eqref{e7} can be thought of in two ways. 

On the one hand it provides a recipe for a laboratory experiment in a microreactor cell. The flow rate $F$, inflow concentrations $c_\text{v,f}$, $c_\text{w,f}$, $c_\text{x,f+y,f}$ and $c_\text{xy,f}$,  inflow temperature $T_\text{f}$ and wall or ambient temperature $T_\text{a}$ are tunable by the experimenter, the wall thermal conductance $L$ and volume $V$ are set by design, and thermokinetic parameters and reaction enthalpies of the RNA reactions can be manipulated by engineering the ribonucleotide sequences and strand lengths. A variety of continuous-flow microreactors are already in use for various calorimetric applications in biotechnology, and the field is developing rapidly \cite{Carreto:2011}. 

On the other hand we need to address the adequacy of the CSTR as a  model for reactions in a porous rock. In an actual physical situation we  assume a flow through  a porous rock structure comprising a large array of individual pores, having patterns of connection determined by the  character of the rock structure. Provided that the system is sufficiently ergodic,  and input flow is steady in volumetric velocity and chemical composition, it is reasonable to assume that the time behaviour in a single pore as reported above gives results at least qualitatively representative of the larger array over times longer than the period of the thermochemical oscillator. 

Further, the use of a CSTR model for a pore implicitly assumes a typical residence time $\gg$ the mixing time. Mixing is achieved mainly by convection, and in this respect thermoconvection times of 3--7.5\,s were  reported  by Mast et al. \cite{Mast:2012}. 
The mean residence time in our system is 20\,s. The perfect mixing assumption is considered to hold adequately when the residence time is around 5--10 $\times $ the mixing time, so the CSTR model can be a reasonably good approximation. The 15\,s thermoconvection time reported by  Braun and Libchaber \cite{Braun:2004}  is too long for the perfect mixing assumption to hold, and in that case the appropriate model is a system of reaction-convection-diffusion equations and boundary conditions. In that case reactive and convective thermal oscillations would couple in important and interesting ways, but that is for future study. We prefer, at this stage, to avoid the sheer `tyranny of numbers' that still plagues the numerical analysis of such systems, and isolate the reactive thermal oscillations by studying a system for which the perfect mixing assumption holds. 

In experiments the mixing behaviour of the system would be  characterised by using a tracer to measure the residence time distribution, or or by determining the period of time  necessary for the system to achieve a desired level of homogeneity. 
We note that Imai et al.  \cite{Imai:1999} confirmed experimentally the well-stirred condition for a  flow microreactor that simulated a hydrothermal environment, in which  elongation of oligopeptides was achieved at high temperature and pressure.

\subsection{RNA stability to hydrogen peroxide}

The response of RNA to H$_2$O$_2$ is a mixed story. Some small RNAs have been found to be very stable; for example a 109-nucleotide RNA  which is induced by  oxidative stress in E. coli has an \textit{in vivo} half-life of 12--30 minutes  \citep{Altuvia:1997}. There is  evidence for modern RNA dysfunction, but not degradation, caused by oxidation initiated by H$_2$O$_2$ \citep{Liu:2012}. In \cite{Cohn:2006} it was found that  small RNA ($\sim$50 bases) of yeast is stable to H$_2$O$_2$, but pyrite is reactive in RNA degradation and hydroxyl (OH) radical generation. 

Much more is known about the effects of H$_2$O$_2$ on DNA, in the context of  disease caused by free radical damage.  In \cite{Seto:1994} it was found that 2$^\prime$deoxyguanosine was hydroxylated at C-8 when included in the thiosulfate-hydrogen peroxide system. But it is an interesting fact that  H$_2$O$_2$ does not react directly with DNA, and DNA is not damaged in its presence unless  transition metal ions are present \citep{Halliwell:1991}.   
In the presence of transition metal ions OH radicals are generated  by the Fenton reaction:
$\text{Fe}^{2+} + \text{H}_2\text{O}_2\rightarrow \text{Fe}^{3+} + \text{OH} + \text{OH}^-.
$
 However, the high and indiscriminate reactivity of the OH radical limits its ability to  damage  biomolecules, because it is more likely to be scavenged by the iron before attacking DNA or RNA:
$\text{OH }+ \text{Fe}^{2+} \rightarrow \text{Fe}^{3+} + \text{OH}^-.
$

Oxidized RNA evidently may still replicate. For example, oxidized mRNA has been successfully converted to cDNA by reverse transcriptase, with mutations induced in the cDNA by the oxidized RNA bases \cite{Kamiya:2007}. 

In summary, although H$_2$O$_2$ may damage or modify RNA (as could many other possible ingredients of primordial soup), there is no reason to suppose that it cannot replicate and pass out of the reaction zone faster than it is damaged or degraded. RNA that is modified by the action of H$_2$O$_2$ in such a way that confers resilience to H$_2$O$_2$ damage would, of course, be selected for. 

  \subsection{Occurrence and supply of substrate RNA oligomers}
Replication of RNA duplexes by complementary strand pairing raises the thorny question of how such strands could have been produced  in the prebiotic primordial soup.  It has been shown that 200-mers of RNA can be polymerized  and concentrated in a thermal gradient \cite{Mast:2013}, but this does not produce complementary strands preferentially. 
Polynucleotides $>$  50-mers can be synthesized on montmorillonite  \cite{Ferris:1996}, which also facilitates homochiral selection \cite{Joshi:2013}, and this would seem a stronger hypothesis. 

\begin{figure}[h]
\centerline{
\includegraphics[scale=1]{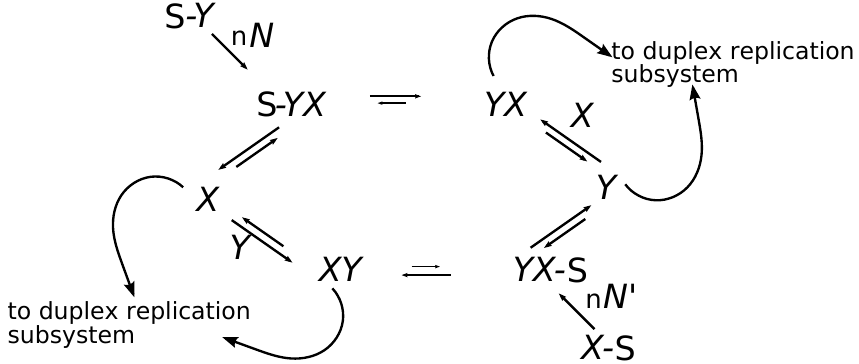}
}
\caption{\label{figure5} Scheme for supply of template oligomers. The surface-bound single polynucleotide strands S-$X$ and S-$Y$  bind n complementary nucleotide  fragments $N$ and $N^\prime$ from solution, which are  ligated irreversibly to form S-$XY$ and S-$YX$.  Heating promotes   release of free  $X$  and $Y$ and slower release of duplex $XY$ from S, and melting of free $XY$ to $X$ and $Y$.  The free species can  participate in the duplex replication scheme (R2) and (R3).   }
% Aspect ratio = 2.35, word count = 84
\end{figure}

A supply of complementary template oligomers could be provided by a variation on the surface-promoted replication procedure  in   \cite{Luther:1998}, when  competitive pathways are included for consumption and production of complementary free RNA species  $X$ and $Y$ as indicated in figure \ref{figure5}. Here, too,  replication  may be driven by  an H$_2$O$_2$ thermochemical oscillator,  because strand separation requires heating  and annealing, ligation, and immobilization of single strands is promoted by cooling. 

 In another pre-biotic scenario, lipid molecules in a dehydrated environment promoted  condensation of nucleic acid monomers, and an alternating wet phase replenished the monomers \cite{Olas:2011}. 
 Such a system also would be enhanced by the H$_2$O$_2$ thermochemical oscillator, which would provide  periodic heating to drive the dehydrations and cooling to allow rehydrations.  

\subsection{Effects of other reactions involving H$_2$O$_2$}
Hydrogen peroxide may undergo other competing reactions; for example, the Fenton reaction in the presence of Fe$^{2+}$/Fe$^{3+}$ discussed above, which is also exothermic. In this case there would be an extra reaction rate term in the species balance for hydrogen peroxide and in the thermal balance. Such a system has an enhanced propensity towards quasiperiodic oscillations.

 It is noteworthy that  the Cu(II)-catalysed  H$_2$O$_2$/thiosulfate ion reaction  displays pH oscillations isothermally  \citep{Orban:1987}, implying that the chemistry alone is sufficiently nonlinear to permit oscillatory dynamics, and  in any case more complex than  (R1) suggests.
  Spontaneous  pH cycling  could well be an assistant driver for RNA replication as it  would provide scope for favourable conformation changes. 

---

In summary, we have proposed and explored a previously unrecognized mechanism for driving a replicating system on the prebiotic earth. The chances of a  hydrogen peroxide thermochemical oscillator arising spontaneously on the early earth in the presence of nucleotide precursors are perhaps very small. However, the earth is large, rock pores are innumerable, and  there was plenty of time on the early earth for improbable events to happen. 

Braun and Libchaber \cite{Braun:2004}  are strong proponents of an interdisciplinary approach between biochemistry and geophysics for understanding the origin of life at the molecular level. We have added some insights that originated in chemical engineering, showing that an interdisciplinary approach is indeed fruitful. 

\bigskip\bigskip
 \begin{table}[h]\caption{\label{table1} Nomenclature and numerical values.} 
\footnotesize\centering{
 \begin{tabular}[b]{>{\PBS\raggedright}p{0.4\columnwidth}p{0.05\columnwidth}cc}
 \hline\hline
Symbol (units), definition &  		Quantity	&				 \multicolumn{2}{c}{Values} \\
  			  &						&  \multicolumn{1}{c}{no RNA} 		     &  \multicolumn{1}{c}{with RNA} \\
\hline\hline
$A$ (M$^{n-1}$s$^{-1}$), pre-exponential  factor&$A_1$ 	&$1.63 \times 10^{10} $&$1.63 \times 10^{10}$\\
		   			&$A_2$ 	&\multicolumn{1}{c}{n/a}				      &$1.69\times 10^{12}$\\
			   &$A_3$ 			&\multicolumn{1}{c}{n/a}				      &$1.0\times 10^{39}$\\
$c$ (M), 		concentration  		 &$c_\text{v,f}$ 		&  0.9			      &  0.9\\
                              &$c_\text{w,f}$ 			&1.35			      &1.35\\	
			   &$c_\text{x,f+y,f}$		&	0			      &$5.0\times 10^{-7}$ \\
			   &$c_\text{xy,f}$ 		 &	0			      &$3.5 \times 10^{-8}$\\
$\bar{C}$ (J\,K$^{-1}$l$^{-1}$), volumetric specific heat&$\bar{C}$ &3400		&3400\\
$E$  (kJ\,mol$^{-1}$), activation energies&$E_1$ 		&68.12			&68.12\\
			&$E_2$ 		& \multicolumn{1}{c}{n/a}				&$-10$\\
			&$E_3$ 	& \multicolumn{1}{c}{n/a}				&260.14\\
$F$ ($\mu$l\,s$^{-1}$), flow rate		&     $F$          &  0.50		&  0.50	\\
$\Delta H$  (kJ\,mol$^{-1}$), reaction	 enthalpies &$\Delta H_1$	&$- 612.5$	& $-612.5$\\
&$\Delta H_2$					&\multicolumn{1}{c}{n/a}				&$-169.6$\\
&$\Delta H_3$ 					&\multicolumn{1}{c}{n/a}				&260.14\\
$L$ (W\,K$^{-1}$), wall thermal conductance& 			&  \\
$n,$ sum of species reaction orders\\
$T$  (K), temperature	&$T_\text{a}$			& 283 & 283\\
$V$ ($\mu$l)  pore volume & 			$V$	& 10.0	 & 10.0\\			
 \hline\hline
  \multicolumn{4}{c}{Subscripts} \\
 \hline\hline
 1, 2, 3 & \multicolumn{3}{l}{pertaining to (R1), (R2), (R3)}\\
 a & \multicolumn{3}{l}{of the ambient or wall temperature}\\
 f  & \multicolumn{3}{l}{of the flow}\\
 v, w & \multicolumn{3}{l}{of thiosulfate, H$_2$O$_2$}\\
  x, y, z & \multicolumn{3}{l}{of single-strands, duplex, quadruplex}\\
  \hline\hline
 \end{tabular}
 }
\end{table} 

\newpage
\section*{Appendix: Choices of numerical values}
\begin{itemize}
\item The weighted average volumetric specific heat $\bar{C}=3400$\,J\,K$^{-1}$l$^{-1}$ is considerably lower than that of pure water.  The volumetric specific heat of sea water is about 3850\,J\,K$^{-1}$l$^{-1}$  and that of H$_2$O$_2$ is about 2620\,J\,K$^{-1}$l$^{-1}$, from which we arrive at the given value of  3400\,J\,K$^{-1}$l$^{-1}$. 
\item The activation energies and pre-exponential factors of the RNA reactions were obtained from the data in \cite{Krammer:2012} using 
$$
E= \frac{R\ln\left(k\left(T_2\right)/k\left(T_1\right)\right)}{1/T_1 - 1/T_2}
$$
then $A=k\exp{(E/RT)}$. The activation energy for the association reaction (R2)  is zero in  \cite{Krammer:2012} but we used the small negative value of 10\,kJ/mol in the calculations to reflect the fact that the rates of association reactions of biological macromolecules often 
decrease with temperature, because due to thermal motion a smaller faction of energetically favourable collisions result in reaction. The data in   \cite{Krammer:2012}  gave a pre-exponential factor for reaction (R3) in the main article of 10$^{45}$\,s$^{-1}$. On the basis of only two data points this should not be taken too literally; in practice we had to reduce it to $10^{39}$\,s$^{-1}$ to couple the reaction to the thermochemical oscillator. 
\end{itemize}

\bigskip\bigskip

\noindent\textit{Acknowledgement: Rowena Ball is recipient of Australian Research Council Future Fellowship FT0991007. }
 
\clearpage 
%\bibliographystyle{elsarticle-num}
% \bibliography{/Users/rowena/bib/Thermalrunaway,/Users/rowena/bib/thermoreplication}

\newpage

\section*{Figure captions and short title for page headings }

\begin{description}
\item[Figure 1.] The quadruplex RNA species $Z$. The duplex replicator species are $XY$ and $xy$, which for the purposes of the simulation are equivalent. See ref. \cite{Krammer:2012} for more details. (Online vesion in colour.)
\item[Figure 2.] Computed data from equations  \eqref{e1}--\eqref{e3}: (a) Hopf (solid  line) and saddle-node (dashed  line) bifurcation loci; (b) Time series for $T_\text{f}=282$\,K, $L=2.4$\,mW;   the oscillation period  is  108.7\,s.  (Online version in colour.)
\item[Figure 3.] Time series from equations \eqref{e1}, \eqref{e2} and \eqref{e4}--\eqref{e7} for $T_\text{f}=282$\,K, $L=2.3$\,mW;  period of the oscillation is  99.3\,s.  (Online version in colour.)
\item[Figure 4.] Onset of oscillations via slow drift of parameters. (a) $p\equiv c_\text{w,f}$, (b) $p\equiv T_\text{a}$, (c) $p\equiv F$, (d) both $p_1\equiv c_\text{w,f}$ and $p_2\equiv T_\text{a}$.  (Online version in colour.)
\item[Figure 5.] Scheme for supply of template oligomers. The surface-bound single polynucleotide strands S-$X$ and S-$Y$  bind n complementary nucleotide  fragments $N$ and $N^\prime$ from solution, which are  ligated irreversibly to form S-$XY$ and S-$YX$.  Heating promotes   release of free  $X$  and $Y$ and slower release of duplex $XY$ from S, and melting of free $XY$ to $X$ and $Y$.  The free species can  participate in the duplex replication scheme (R2) and (R3).
\item[Short title:] Primordial RNA replication
\end{description}
\end{document}